\documentclass[rnote,bibyear]{aa}  

\usepackage{graphicx}
\usepackage{txfonts}
\begin{document}

   \title{Overlooked wide companions of nearby F stars}


\titlerunning{Overlooked wide companions}

   \author{R.-D. Scholz}

   \institute{Leibniz-Institut f\"ur Astrophysik Potsdam (AIP),
              An der Sternwarte 16, 14482 Potsdam, Germany\\
              \email{rdscholz@aip.de}
             }

   \date{Received 15 December 2015; accepted 7 January 2016 }

 
  \abstract
   {}
   {We checked a sample of 545 F stars within 50~pc 
for wide companions using existing near-infrared and 
optical sky surveys.
}
   {Applying the common proper motion (CPM) criterion, 
we detected wide companion candidates with 6-120~arcsec 
angular separations by visual inspection of multi-epoch 
finder charts and by searching in proper motion 
catalogues. Final proper motions were measured by 
involving positional measurements from up to eleven 
surveys. Spectral types of red CPM companions were 
estimated from their absolute $J$-band magnitudes 
based on the Hipparcos distances of the primaries.
}
   {In addition to about 100 known CPM objects,
we found 19 new CPM companions and confirmed 31 
previously known candidates. A few CPM objects are still 
considered as candidates according to their level of 
proper motion agreement. Among the new objects there are 
nine M0-M4, eight M5-M6, one $\approx$L3.5 dwarf (HD\,3861B), 
and one white dwarf (WD) (HD\,2726B), 
whereas we confirmed two K, 19 M0-M4, 
six M5-M6, 
two early-L dwarfs, and two DA WDs as CPM 
companions. In a few cases, previous spectral types were 
available that all agree well with our estimates. Two 
companions (HD\,22879B and HD\,49933B) are associated with
moderately metal-poor Gaia benchmark stars. One doubtful 
CPM companion, spectroscopically classified as WD but 
found to be very bright ($J$$=$11.1) by others, should 
either be a very nearby foreground WD or a different kind 
of object associated with HD\,165670.
}
   {}


   \maketitle
%

\section{Introduction}
\label{Sect_intro}

Stellar binaries and multiple systems appear to be the main pro\-duct of star 
formation, if the primaries of these systems are about at least as massive
as the Sun. The solar neighbourhood, representing the average Galactic disk 
population of stellar systems that exist already for typically several Gyrs, 
is naturally one of the best-investigated regions
with respect to multiplicity. The Research Consortium on Nearby Stars
(RECONS)\footnote{http://www.chara.gsu.edu/RECONS/} states a multiplicity
rate, i.e. the probability that a given system has more than one component, 
of 29\% for their 10~pc sample. However, this overall ''low rate is because 
M-type dwarfs dominate the solar neighbourhood (a full 73\% of the stellar 
sample ... are M-type dwarfs), and do not have companions as often as their 
more massive stellar cousins''. Among the 100 nearest RECONS systems (with 
a horizon of about 6.5~pc), there are 70 with M dwarfs as the most-massive
component with only 18 (26\%) known to have companions. On the other hand,
there are 22 AFGK primaries with 12 of them (55\%) in known multiple systems.
The RECONS 10~pc census shows a strong increase from 2000 to 2012 both
in the number of M dwarfs ($+$25\%) and in the number of stellar and
LT-type companions ($+$26\%). 

The multiplicity of F- and G-type stars in a wider solar neighbourhood
was in the focus of investigations by Fuhrmann \& Chini~(\cite{fuhrmann12}, 
\cite{fuhrmann15}), Chini et al.~(\cite{chini14}), 
Tokovinin~(\cite{tokovinin11}, \cite{tokovinin14}), and Tokovinin \&
L{\'e}pine~(\cite{tokovinin12}). To distinguish between visual and
physical wide double stars, the common proper motion (CPM) of the
components is often used as a criterion. In the era of photographic
sky surveys, Luyten~(\cite{luyten97}) compiled a catalogue of
CPM pairs.
Frankowski et al.~(\cite{frankowski07}) investigated CPM binaries in 
the Hipparcos catalogue, with both components being Hipparcos stars,
compared their short-term (Hipparcos) and long-term (Tycho-2) proper motions
and used radial velocities as control data. The systematic search for faint 
CPM companions to Hipparcos stars (Gould \& Chanam{\'e}~\cite{gould04}),
L{\'e}pine \& Bongiorno~\cite{lepine07}) not only improved the statistics
of the multiplicity of AFGK stars, but also allowed for a better 
characterisation of the lower-mass companions by making use of the 
knowledge about the primaries. One of the physical
parameters of interest in that respect is the metallicity of M dwarfs 
and subdwarfs (e.g. Li et al.~\cite{li14}). The CPM method continues
to be useful, in particular for new deep surveys 
(e.g. Ivanov et al.~\cite{ivanov13}, Deacon et al.~\cite{deacon14}).
Focusing on a sample of nearby F stars,
we demonstrate how much we can still improve the CPM statistics
and find previously overlooked wide companions, even of well-known 
bright stars, in existing public surveys. The main results
of this research note, data on new, confirmed, and rejected 
CPM companions, are listed in
Tables~\ref{Tab_newcpmdata}, \ref{Tab_confirmedcpmdata}, 
and \ref{Tab_rejectedcpmdata}, respectively.

%

\section{Sample definition and search method}
\label{Sect_sample}

\subsection{SIMBAD F stars sample}
\label{SubSect_SIMBAD}

From 1188 stars, for which SIMBAD lists spectral types F0 to F9 
and parallaxes larger than 20~mas, we selected 545 with
proper motions larger than 150~mas/yr to search for CPM
companions. Our high proper motion 50~pc sample overlaps 
in part with the about nine times
larger sample of Tokovinin~(\cite{tokovinin14}). His
sample reaches out to 67~pc, excludes stars with 
large parallax errors ($>$7~mas) and contains both F and G 
stars, but with a colour selection corresponding to only
late-F and early-G stars (approximately F5V to G6V).
Finally, he included only dwarfs and sub-giants.
We are aware of possible uncertainties
(missing updates) of the SIMBAD data, in particular the
spectral types (see also Scholz et al.
\cite{scholz15} concerning F-type subdwarfs). 

\subsection{Visual inspection and catalogue search}
\label{SubSect_viscat}

In our visual inspection of the sky areas around 
545 F stars, we used
IRSA finder charts tools\footnote{old:
http://irsa.ipac.caltech.edu/applications/FinderChart/ 
(for DSS, SDSS, and 2MASS 1-6~arcmin image sizes),
new:
http://irsa.ipac.caltech.edu/applications/finderchart/
(with additional WISE images, and allowing for smaller
image sizes)}
in several runs with image sizes from 0.5~arcmin to 3~arcmin. 
In addition, we extracted 1~arcmin and 2~arcmin finder charts
from the UKIRT deep infrared sky survey (UKIDSS) and the visible 
and infrared survey telescope for astronomy (VISTA) archives\footnote{
http://wsa.roe.ac.uk/ and http://horus.roe.ac.uk/vsa/}.
We considered the two micron all-sky survey 
(2MASS; Skrutskie et al.~\cite{skrutskie06}) images with their
high resolution and dynamic range as reference and searched for
CPM candidates in other images with epoch differences large enough 
to show a change in position similar to that of the primaries.
Special attention was paid to objects overlapping with diffraction
spikes and other known artefacts, as well as to those with very red 
or blue colours expected for very low-mass and white dwarf (WD) companions,
respectively.
A few cases are illustrated by selected finder charts in 
Figs.~\ref{F_HD22879AB}, \ref{F_HD3861AB}, and \ref{F_HD2726AB}.

Our search was sensitive to angular separations from a few arcseconds
to a few arcminutes. It was complementary to the work of 
Tokovinin \& L{\'e}pine~(\cite{tokovinin12}) that aimed at
angular separations 
of the companion larger than 30\,arcsec. We tried
to find CPM companions as close as possible with the existing sky 
surveys. However, to detect very close companions with 
angular separations
of the order of 1~arcsec or less, dedicated high-resolution
imaging observations (e.g. Ehrenreich et al.~\cite{ehrenreich10},
Meshkat et al.~\cite{meshkat15}) 
are required.

Using the CDS cross-match service\footnote{http://cdsxmatch.u-strasbg.fr/xmatch}
we also checked the fourth US Naval Observatory CCD astrograph catalogue 
(UCAC4; Zacharias et al.~\cite{zacharias13}) and the first US Naval 
Observatory robotic astrometric telescope catalogue
(URAT1; Zacharias et al.~\cite{zacharias15}) for possible new CPM 
companions within 2~arcmin of the primaries. Note that these catalogues were
already subject of CPM searches by Hartkopf et al.~(\cite{hartkopf13})
and Nicholson~(\cite{nicholson15}). In
our catalogue search, we did not require a correct proper motion measurement
of the bright and sometimes problematic primaries in the given catalogue.

\subsection{Proper motion measurements and CPM status}
\label{SubSect_pmfi}

In our proper motion determinations we combined multi-epoch
positional measurements from the following surveys (roughly sorted by epochs) 
if available:\\
\textbf{\scriptsize 
1) photographic Schmidt plates scanned with the
   automated photographic measuring (APM; McMahon, Irwin, \& Maddox~\cite{mcmahon00})
   and SuperCOSMOS sky surveys (SSS; Hambly et al.~\cite{hambly01}),\\
2) 2MASS (Skrutskie et al.~\cite{skrutskie06}),  \\
3) deep near-infrared southern sky survey (DENIS; Epchtein et al.~\cite{epchtein97}),\\
4) UCAC4 (Zacharias et al.~\cite{zacharias13}),\\
5) Carlsberg meridian catalogue (CMC; Mui{\~n}os \& Evans~\cite{muinos14}), \\
6) Sloan digital sky survey (SDSS; Abazajian et al.~\cite{abazajian09}),\\
7) INT photometric H$\alpha$ survey (IPHAS; Barentsen et al.~\cite{barentsen14}),\\
8) UKIDSS large area 
   (UKIDSS LAS; Lawrence et al.~\cite{lawrence07})
   and galactic plane surveys (UKIDSS GPS; Lucas et al~\cite{lucas08}), \\
9) wide-field infrared survey explorer (WISE; Wright et al.~\cite{wright10}), \\
10) URAT1 (Zacharias et al.~\cite{zacharias15}),\\
11) VISTA hemisphere (VHS; McMahon et al.~\cite{mcmahon13}) and 
Variables in the Via Lactea surveys (VVV; Minniti et al.~\cite{minniti10}).\\}
Those CPM candidates with small separations that were not well-measured or
absent in the SSS/2MASS/DENIS/SDSS catalogues, we detected visually
in the corresponding FITS images using the ESO skycat tool.
Depending on the number of epochs available, the accuracy of the simple
linear proper motion fit that used all input positions with equal weights
varied considerably. No attempt was made to transform the target 
positions in different surveys to a common system before the proper motion 
fit, as we expected individual centroiding errors affected by the close
bright primaries to be larger ($\gtrapprox$100~mas) than systematic errors. 

The proper motion errors of the CPM companions were typically much
larger than those of the known primaries, although in some cases we 
achieved a high precision for the CPM companion and excellent agreement 
with the known proper motion of its primary. For the majority of the
19 new and 31 confirmed CPM companions shown in Tables~\ref{Tab_newcpmdata}
and \ref{Tab_confirmedcpmdata}, respectively, their proper motion components agreed to 
within 2~$\sigma$ of the formal errors with those of the primaries. For the primaries,
we preferred the longer-time baseline proper motions of the UrHip (Frouard et al.~\cite{frouard15})
or Tycho-2 (H{\o}g et al.~\cite{hog00}) catalogues instead of the Hipparcos 
(van Leeuwen~\cite{vanleeuwen07}) values. A few CPM companions, for which
at least one of the proper motion components agreed only to within 3~$\sigma$
with that of the primary, are marked by (?) as CPM candidates, whereas
two objects with even larger discrepancies are considered as doubtful (??)
CPM objects. Our overall level of agreement of the proper motion components is
similar to that of Deacon et al.~(\cite{deacon14}), who described 57 new faint
CPM objects of Hipparcos stars with total proper motion differences of less than 5~$\sigma$ (their Eq.~1).
 
\subsection{Spectral type estimates}
\label{SubSect_SpT}

The spectral types of red (according to their near-infrared 2MASS, UKIDSS or
VISTA colours, and DENIS $I$$-$$J$ colours if available)
CPM companions were estimated based on the
known distances of the primaries and the relation between absolute
$J$ magnitude and (early-K to late-L) spectral type from
Scholz, Meusinger \& Jahrei{\ss}~(\cite{scholz05}). Our
spectral type estimates are in good agreement (within 0.5 subtypes)
with previous classifications available for six M and two early-L 
dwarfs (Tables~\ref{Tab_newcpmdata} and \ref{Tab_confirmedcpmdata}).
Blue objects, simply assumed to be WD candidates, as well as
possible subdwarf candidates are discussed in Sect.~\ref{Sect_discussion}.

%

\section{Newly found CPM companions}
\label{Sect_newcpm}


Table~\ref{Tab_newcpmdata} presents our new CPM discoveries.
Here we include some CPM companions with previous proper motion 
measurements (three in UCAC4, four in URAT1) and/or spectral 
classifications (one early-M dwarf), whose association 
with the primary was not mentiond before. All new CPM objects
belong to primaries with distances between 25 and 50~pc according 
to their Hipparcos parallaxes.
One object (HD\,76493B) is a doubtful CPM companion not only
because of the poor agreement of $\mu_{\alpha}\cos{\delta}$,
but also because of the discrepancy between the short- and
long-term proper motions of its primary HD\,76493A. Such 
discrepancies hinting at the influence of unresolved companions 
are also seen for the known close binary HD\,2057AB
and for several other primaries in both 
Tables~\ref{Tab_newcpmdata} and \ref{Tab_confirmedcpmdata}.
If confirmed as CPM, HD\,76493B would have the largest projected 
physical separation of all our targets ($\approx$2600~AU).

%

\section{Brief discussion and notes on individual objects}
\label{Sect_discussion}

In addition to the objects listed in Tables \ref{Tab_newcpmdata} and \ref{Tab_confirmedcpmdata}
there were 98 other known CPM objects among the 545 F stars (18\%),
mainly at small ($<$5~arcsec) or large ($>$120~arcsec)
angular separations and of earlier spectral types (FGK). With our 19$+$31
new and confirmed CPM companions we add further 9\% (50\% more
CPM objects!), which
are mainly M dwarfs at intermediate angular separations. Both
new and confirmed objects have on average projected physical separations 
of about 900~AU (ranging from about 200~AU to 2500~AU).
Certainly, the high resolution astrometric measurements of Gaia
will bring the multiplicity rate of the F stars in our high
proper motion 50~pc sample to the 50\% level, as currently known 
for the small RECONS 6.5~pc sample, 
or even higher. Gaia will also provide accurate
distances for individual system components and the CPM status for all 
nearby objects including those with relatively small proper motions.

We checked our primaries for clearly (repeatedly measured) non-solar
metallicities using VizieR and found only two metal-poor stars, HD\,22879A and
HD\,49933A among our new and confirmed CPM systems, respectively.
Both are Gaia benchmark stars with respect to metallicity 
(Jofr{\'e} et al.~\cite{jofre14}) with mean literature [Fe/H] values
of $-$0.85 and $-$0.39, respectively. Their CPM companions,
HD\,22879B and HD\,49933B, are therefore M subdwarf candidates
that can be used for the calibration of M dwarf metallicities
(e.g.  Neves et al.~\cite{neves12}, Newton et al.~\cite{newton14}).

Our new and confirmed wide CPM companions of nearby F stars 
distributed all over the sky represent good targets for spectroscopic 
follow-up observations, to verify their spectral types and 
to confirm their physical association with the
primaries by radial velocity measurements. 
Our lowest-mass new CPM companion, the suspected
$\approx$L3.5 dwarf HD\,3861B, is of particular interest, as
L-type companions of nearby F-type stars are rare 
(Wilson et al.~\cite{wilson01}, Luhman et al.~\cite{luhman12}, 
Gauza et al.~\cite{gauza12},
Deacon et al.~\cite{deacon14}). Our spectral type estimate
for HD\,3861B is also supported by its $J$$-$$K$=$+$1.7 measured
in the UKIDSS LAS,
which is a typical colour of a mid-L dwarf (Leggett et al.~\cite{leggett10}).

The CPM criterion is also used for membership probability in 
moving star clusters (e.g. Gagn{\'e} et al.~\cite{gagne15}). One
of our new CPM companions, HD\,175317B (Table~\ref{Tab_newcpmdata}), 
was previously considered as AB Dor moving group member by
Malo et al.~(\cite{malo13}), who did not mention the small separation
($\approx$18~arcsec) and CPM with respect to HD\,175317A. We consider 
this relatively close CPM pair as most likely 
physically bound, although the $\mu_{\alpha}\cos{\delta}$ agreement
is only within 3~$\sigma$. This does not exclude a moving group membership.
The confirmed CPM companion HD\,126679B (Table~\ref{Tab_confirmedcpmdata}) 
was investigated by Gagn{\'e} et al.~(\cite{gagne15}) but not found to 
be a member in any moving group.

A strong decline in the frequency of Sirius-like systems (AFGK stars
with WD companions) beyond a distance of 20~pc was mentioned
by Holberg et al.~(\cite{holberg13}), who predicted new discoveries
of such systems with different observing techniques. In our search,
we confirmed two WD CPM companions 
(Table~\ref{Tab_confirmedcpmdata}) and found one 
previously overlooked at a separation of 
about 
20~arcsec to the early-F
star HD\,2726A (Table~\ref{Tab_newcpmdata}). 
The new object, HD\,2726B, was hardly seen in 2MASS
but well-detected in DENIS ($J$$=$14.96, $I$$-$$J$$=$$-$0.33) and 
VISTA VHS (mean $J$$=$14.83, $J$$-$$K$$=$$-$0.19). Both its proper 
motion components agree to within 1~$\sigma$ with those of HD\,2726A.

Our most doubtful CPM confirmation, HD\,165670H, is not red enough
for an early-M dwarf classification 
(2MASS $J$$=$10.27, $J$$-$$K_s$$=$$+$0.39) and was therefore also
considered as WD candidate. Deacon et al.~(\cite{deacon14}) classified
it as DA WD based on a near-infrared spectrum and found a fainter
magnitude and bluer colour (UKIRT $J$$=$11.10, $J$$-$$K$$=$$+$0.15). Their
measured separation of 9.6~arcsec is 
1.8~arcsec larger than ours, indicating
centroiding problems or a change over time (no CPM?),
but their proper motion ($+$36$\pm$5, $-$133$\pm$5) is in better agreement
with that of the primary. However, the primary shows a discrepancy between 
its Hipparcos and UrHip proper motions. More importantly, the $J$
magnitude of HD\,165670H is comparable to that of the nearest known
WDs (Fig.~1 in Scholz et al.~\cite{scholz15}). Therefore, this is
either a very nearby WD in the foreground
or a different kind of object (hot subdwarf?) associated with
HD\,165670A.

%
%
\begin{table*}
\caption{Data on nearby F stars and their new CPM companions (Possible and doubtful CPM candidates are marked with (?) and (??), respectively)}
\label{Tab_newcpmdata}
\centering
\fontsize{7pt}{0.90\baselineskip}\selectfont
\begin{tabular}{@{}lrrc@{}c@{}cccccr@{}}     
\hline\hline
Name & RA (J2000) & DEC (J2000) & Epoch  & $\pi$ (Sep, PA)    & $\mu_{\alpha}\cos{\delta}$ & $\mu_{\delta}$ & N\_epochs & $J$ (2MASS) &SpT & Ref \\
HD     & [degrees]  & [degrees]   & [year] & [mas] ([arcsec,$^{\circ}$])& [mas/yr]                   & [mas/yr]      & & [mag]       & \\
\hline
2057AB   &   6.286235 & $+$48.047478 & 1998.849 & 22.82$\pm$0.88 & $+$281.3$\pm$0.4\tablefootmark{j} &   $+$1.0$\pm$0.4\tablefootmark{j} & & 6.38$\pm$0.02 & F8 & 2,3,4,S\\
2057D\tablefootmark{s}   &   6.280679\tablefootmark{a} & $+$48.046743\tablefootmark{a} & 1998.849 & (13.6, 259)         & $+$283$\pm$26\tablefootmark{a,c}    &   $-$3$\pm$10\tablefootmark{a,c}    & 11000000010 & $\approx$12.5\tablefootmark{e} & $\approx$M5 & 2,2,5,5\\
\hline
2726A   & 7.608672 & $-$48.214951 & 1999.767 & 23.11$\pm$0.43 & $+$134.2$\pm$0.8\tablefootmark{j} & $-$84.5$\pm$0.8\tablefootmark{j} & & 4.94$\pm$0.04\tablefootmark{f} & F2V & 2,3,8,S \\
2726B   & 7.600967\tablefootmark{a} & $-$48.212594\tablefootmark{a} & 1999.767 & (20.3, 295)         & $+$135$\pm$12\tablefootmark{a,c}    & $-$94$\pm$13\tablefootmark{a,c}    & 21110000001 & 14.96$\pm$0.14\tablefootmark{g} & WD & 2,2,5,5 \\
\hline
3861A  & 10.2991063 & $+$9.3548251 & 2010.592 & 29.90$\pm$0.45 & $-$125.0$\pm$0.6\tablefootmark{j} & $-$103.7$\pm$0.6 & & 5.58$\pm$0.03 & F8V & 7,3,4,S \\
3861B  & 10.2945415 & $+$9.3556005 & 2010.592 & (16.5, 280) & $-$121$\pm$14\tablefootmark{a,d} & $-$79$\pm$13\tablefootmark{a,d} & 01000102000 & 15.690$\pm$0.006\tablefootmark{g} & $\approx$L3.5  & 7,7,5,5 \\
\hline
10226A  & 24.977409  & $-$9.972228  & 1998.816 & 20.35$\pm$0.81 & $+$239.8$\pm$1.7 & $+$92.2$\pm$1.3 & & 6.73$\pm$0.02 & F8  & 2,3,8,S \\
10226B  & 24.975580  & $-$9.973311  & 1998.816 & (7.6, 239)          & $+$229$\pm$21\tablefootmark{d}    & $+$76$\pm$16\tablefootmark{d}    & 01000200000 & $\approx$12.6\tablefootmark{e} & $\approx$M5 & 2,2,5,5 \\
\hline
18404A   & 44.521691 & $+$20.668737 & 1998.751 & 30.15$\pm$0.30 & $+$235.4$\pm$0.6 & $-$29.8$\pm$0.6 & & 5.31$\pm$0.20 & F5IV & 2,3,4,S \\
18404B   & 44.525723 & $+$20.667135 & 1998.751 & (14.8, 113)         & $+$232$\pm$7     & $-$21$\pm$4     & 01010100110 & 10.47$\pm$0.07& $\approx$M3.5 & 2,2,5,5 \\
\hline
22879A  & 55.091754  & $-$3.216831  & 1998.732 & 39.12$\pm$0.56 & $+$688.6$\pm$0.6 & $-$213.0$\pm$0.6 & & 5.59$\pm$0.02 & F9V & 2,3,4,S \\
22879B  & 55.086450\tablefootmark{a}  & $-$3.217122\tablefootmark{a}  & 1998.732 & (19.1, 267)         & $+$672$\pm$15\tablefootmark{a,c}    & $-$231$\pm$14\tablefootmark{a,c}    & 11100000010 & 12.09$\pm$0.07\tablefootmark{g} & $\approx$M6/sdM? & 2,2,5,5 \\
\hline
35681A   & 82.003845 & $+$33.763836 & 1998.090 & 29.54$\pm$0.51 & $+$18.2$\pm$0.4 & $-$211.0$\pm$0.4 &  & 5.46$\pm$0.02 & F7V & 2,3,4,S \\
35681E\tablefootmark{s} (?)  & 82.004011 & $+$33.761101 & 1998.090 & (9.9, 177)          & $+$33$\pm$8     & $-$231$\pm$7     & 01000011010 & 10.32$\pm$0.08 & $\approx$M3.5 & 2,2,5,5 \\
\hline
57334A  & 109.551348 & $-$51.050465 & 2000.195 & 20.27$\pm$0.52 & $+$24.2$\pm$1.0 & $-$150.5$\pm$1.0 & & 6.49$\pm$0.02 & F9V & 2,3,8,S  \\
57334B  & 109.543129 & $-$51.059307 & 2000.195 & (36.9, 210)         & $+$20$\pm$6\tablefootmark{b,i}     & $-$158$\pm$4\tablefootmark{b,i}     & 61110000100 & 11.38$\pm$0.02 & $\approx$M3.5 & 2,2,5,5  \\
\hline
74868A   & 131.211449  & $-$44.542557 & 2001.090 & 27.41$\pm$0.36 & $-$194.0$\pm$1.2 & $+$135.4$\pm$1.0 & & 5.63$\pm$0.02 & F9 & 2,3,8,S \\
74868B (?)  & 131.215346\tablefootmark{a}  & $-$44.542417\tablefootmark{a} & 2001.090 & (10.0, 87)          & $-$231$\pm$15\tablefootmark{a,b}    & $+$158$\pm$17\tablefootmark{a,b}    & 01100000000 & $\approx$11.9\tablefootmark{e} & $\approx$M5 & 2,2,5,5 \\
\hline
76493A   & 134.444037 & $+$38.258907 & 2000.022 & 21.52$\pm$1.58 &  $-$119$\pm$0.4\tablefootmark{j} & $-$87.6$\pm$0.4\tablefootmark{j} &  & 6.13$\pm$0.03 & F5 & 2,3,4,S \\
76493B (??)  & 134.463382 & $+$38.256172 & 2000.022 & (55.6, 100)      &  $-$98$\pm$5\tablefootmark{h}    & $-$79$\pm$5\tablefootmark{h}     & 61011200100 & 13.09$\pm$0.03 & $\approx$M5.5 & 2,2,5,5 \\
\hline
129926A & 221.500396 & $-$25.443171 & 1998.479 & 33.02$\pm$0.92 & $-$151.5$\pm$1.1 & $-$107.3$\pm$0.6 & & 4.12$\pm$0.63\tablefootmark{f} & F0IV & 2,3,3,S   \\
129926D\tablefootmark{s} & 221.504310 & $-$25.449644 & 1998.479 & (26.6, 151) & $-$145$\pm$16\tablefootmark{i} & $-$123$\pm$18\tablefootmark{i} & 01100000200 & 10.90$\pm$0.03 & $\approx$M4 & 2,2,5,5 \\
\hline
141103A & 236.820746 & $-$0.269923 & 2006.568 & 20.55$\pm$0.64 & $-$243.0$\pm$0.3 & $-$21.3$\pm$0.3 &  & 5.94$\pm$0.02 & F5 & 7,3,4,S \\
141103B & 236.816820 & $-$0.269806 & 2006.568 & (14.1, 272)         & $-$243$\pm$16\tablefootmark{a,d}    & $-$12$\pm$39\tablefootmark{a,d}    & 01000101000 & $\approx$13.5\tablefootmark{e} & $\approx$M6 & 7,7,5,5 \\
\hline
143306A & 240.618486 & $-$56.854485 & 2000.170 & 29.56$\pm$0.40 & $-$148.3$\pm$4.4 & $-$117.1$\pm$4.1 & & 5.60$\pm$0.03 & F8V & 2,3,8,S \\
143306B & 240.638832 & $-$56.849342 & 2000.170 & (44.1, 65)         & $-$146$\pm$3\tablefootmark{c,i}     & $-$118$\pm$4\tablefootmark{c,i}     & 41210000100 & 10.46$\pm$0.02 & $\approx$M3.5 & 2,2,5,5  \\
\hline
155060A & 256.9818847 & $+$32.1053256 & 2013.918 & 27.36$\pm$0.43 & $-$161.7$\pm$0.3 & $-$41.4$\pm$0.3 & & 6.12$\pm$0.03 & F8 & 1,3,4,S \\
155060B & 256.9812906 & $+$32.1036278 & 2013.776 & (6.4, 197)          & $-$170$\pm$9\tablefootmark{h}     & $-$49$\pm$13\tablefootmark{h}    & 01000200010 & 9.28$\pm$0.17 & $\approx$M1 & 1,1,5,5 \\
\hline
175317A & 283.879186 & $-$16.376575 & 1998.290 & 31.53$\pm$0.33 & $-$27.9$\pm$0.7 & $-$184.4$\pm$0.7 & & 4.73$\pm$0.04\tablefootmark{f} & F5V & 2,3,8,S \\
175317B\tablefootmark{p} (?) & 283.882362 & $-$16.380438 & 1998.290 & (17.7, 142)         & $-$37$\pm$4\tablefootmark{h}     & $-$180$\pm$3\tablefootmark{h}     & 11110000110 & 9.13$\pm$0.03 & $\approx$M1\tablefootmark{q} & 2,2,5,5 \\
\hline
181096A & 289.2140625 & $+$47.0002178 & 2013.810 & 23.79$\pm$0.32 & $-$8.9$\pm$0.4 & $+$291.2$\pm$0.4 & & 5.15$\pm$0.27\tablefootmark{f} & F6IV: & 1,3,4,S \\
181096B & 289.2041292 & $+$47.0092125 & 2013.661 & (40.5, 323)         & $-$9$\pm$3\tablefootmark{c} & $+$289$\pm$1\tablefootmark{c} & 41010000110 & 10.09$\pm$0.02 & $\approx$M2.5 & 1,1,5,5 \\
\hline
188769A & 300.253517 & $-$64.807938 & 2000.430 & 24.11$\pm$0.64 & $+$154.3$\pm$1.3\tablefootmark{j} & $-$242.1$\pm$1.3\tablefootmark{j} & & 6.01$\pm$0.02 & F3IV & 2,3,8,S \\
188769B & 300.238788\tablefootmark{a} & $-$64.807814\tablefootmark{a} & 2000.430 & (22.6, 271)\tablefootmark{a}         & $+$155$\pm$17\tablefootmark{a,c}    & $-$194$\pm$38\tablefootmark{a,c}    & 11200000000 & $\approx$12.6\tablefootmark{e} & $\approx$M5 & 2,2,5,5 \\
\hline
215588A & 341.2647036 & $+$58.1465458 & 2013.438 & 28.22$\pm$0.32 & $-$61.0$\pm$0.4 & $-$135.9$\pm$0.4 & & 5.95$\pm$0.03 & F5 & 1,3,4,S  \\
215588B & 341.2647767 & $+$58.1425542 & 2013.468 & (14.4, 179)         & $-$56$\pm$4\tablefootmark{h}     & $-$138$\pm$4\tablefootmark{h}     & 01010011110 & 9.97$\pm$0.05 & $\approx$M3 & 1,1,5,5  \\
\hline
218235A & 346.575431 & $+$18.517733 & 1997.773 & 22.96$\pm$0.49 & $+$227.7$\pm$1.1 & $+$60.3$\pm$1.0 & & 5.30$\pm$0.02 & F6Vs & 2,3,8,S  \\
218235B & 346.570271\tablefootmark{a} & $+$18.518478\tablefootmark{a} & 1997.773 & (17.8, 279)\tablefootmark{a} & $+$229$\pm$22\tablefootmark{a,d} & $+$69$\pm$7\tablefootmark{a,d} & 01000100000 & $\approx$12.8\tablefootmark{e} & $\approx$M5.5 & 2,2,5,5 \\
\hline
\end{tabular}
\tablefoot{\fontsize{7pt}{0.90\baselineskip}\selectfont
The 11-digit number N\_epochs gives the number of epochs from SSS/APM, 
2MASS, DENIS, UCAC4, CMC, SDSS, IPHAS, UKIDSS, WISE, URAT1, and VISTA, which were used
in the proper motion determination. Visual position measurements in images of:
\tablefoottext{a}{2MASS,}
\tablefoottext{b}{DENIS,}
\tablefoottext{c}{SSS,}
\tablefoottext{d}{SDSS.}
\tablefoottext{e}{Not measured by 2MASS, estimated from visual comparison with other 2MASS objects in the field,} 
\tablefoottext{f}{poor photometry according to 2MASS quality flag,}
\tablefoottext{g}{$J$ magnitude from DENIS or UKIDSS.}
\tablefoottext{h}{Similar proper motion in URAT1 (Zacharias et al.~\cite{zacharias15}),}
\tablefoottext{i}{similar proper motion in UCAC4 (Zacharias et al.~\cite{zacharias13}).}
\tablefoottext{j}{Hipparcos proper motion of van Leeuwen~(\cite{vanleeuwen07}) and
UrHip or Tycho-2 proper motion do not agree within their errors.}
Listed (without or with a different proper motion) in:
\tablefoottext{k}{Tokovinin~(\cite{tokovinin14}) or Tokovinin~(\cite{tokovinin11}),}
\tablefoottext{l}{WDS (Mason et al.~\cite{mason01}).}
\tablefoottext{m}{Also in Wycoff et al.~(\cite{wycoff06}).}
\tablefoottext{n}{Discovered by Mugrauer et al.~(\cite{mugrauer04}).}
\tablefoottext{o}{Also in Hartkopf et al.~(\cite{hartkopf13}).}
\tablefoottext{p}{Member of AB\,Dor moving group according to Malo et al.~(\cite{malo13}).}
\tablefoottext{q}{Riaz et al.~(\cite{riaz06}) determined M0.5 spectroscopically.}
\tablefoottext{r}{Mentioned as CPM by Nicholson~(\cite{nicholson15}).}
\tablefoottext{s}{Other (visual or physical) components are listed in the WDS.}
\tablefoottext{t}{Discovered by Gauza et al.~(\cite{gauza12}), who classified it
spectroscopically as L1. Components B and C form a close binary system composed of an
M8 and L3 dwarf (Gizis et al.~\cite{gizis03}) found to be co-moving with HD\,221356A
at a very wide separation of about 452~arcsec (Caballero~\cite{caballero07}).} 
\tablefoottext{u}{Discovered by Wilson et al.~(\cite{wilson01}) with 
spectroscopic classification of L0.}
\tablefoottext{v}{Similar proper motion in Naval Observatory 
merged astrometric dataset (NOMAD; Zacharias et al.~\cite{zacharias04}) with 
reference to an unpublished YB6 catalogue.}
\tablefoottext{w}{Discovered by Luhman et al.~(\cite{luhman12}) and classified as M2.}
\tablefoottext{x}{Sep=85~arcsec (?) in Luyten~(\cite{luyten97}).} 
\tablefoottext{y}{Discovered and classified as M4.5 by Lowrance et al.~(\cite{lowrance02}).}
\tablefoottext{z}{SIMBAD lists an identical proper motion as for the primary without reference
and a spectral type of M1 according to Bidelman~(\cite{bidelman85}).}
\tablefoottext{$\otimes$}{Discovered and classified as DA WD CPM companion Hip\,88728B by Deacon et al.~(\cite{deacon14})}
\tablefoottext{$\dagger$}{Shares a common radial velocity with the primary 
according to RAVE (Kordopatis et al.~\cite{kordopatis13}).}
\tablefoottext{$\ddagger$}{According to Gagn{\'e} et al.~(\cite{gagne15}) a nearby potential $>$M5 dwarf.}
\tablefoottext{$\star$}{Discovered by Chini et al.~(\cite{chini14}).}
\tablefoottext{$\ast$}{Bidelman~(\cite{bidelman80}) mentioned as wide CPM companion 
and estimated a spectral type of M2 or M3.}
\tablefoottext{$\bullet$}{Exoplanetary host star.}
\tablefoottext{$\clubsuit$}{SIMBAD lists 
a  Hipparcos parallax 
of 47.8$\pm$5.0~mas
and proper motion similar to that of Hip\,10531.
Comparison of SSS, 2MASS and SDSS finder charts confirms the small URAT1
proper motion indicating that Hip\,10529 is
a background object unrelated to the nearby star Hip\,10531.}
\tablefoottext{$\spadesuit$}{SIMBAD erroneously lists a large proper motion similar to that of HD\,187691A 
and spectral type of M4, which should be assigned to HD\,187691C
separated by 
22~arcsec.}
}
\tablebib{\fontsize{7pt}{0.90\baselineskip}\selectfont
Position/Epoch, parallax of primary (angular separation 
and position angle
of CPM companion
- for URAT1 data, we neglected small epoch differences of $\lessapprox$0.3~years), 
proper motion, spectral type: 
(S) SIMBAD;
(1) URAT1;
(2) 2MASS; 
(3) Hipparcos (van Leeuwen~\cite{vanleeuwen07});
(4) UrHip (Frouard et al.~\cite{frouard15});
(5) this paper;
(6) UKIDSS GPS;
(7) UKIDSS LAS;
(8) Tycho-2 (H{\o}g et al.~\cite{hog00});
(9) corrected Hipparcos data (Fabricius \& Makarov~\cite{fabricius00});
(10) SPM4 (Girard et al.~(\cite{girard11});
(11) Limoges et al.~(\cite{limoges13}).
}
\end{table*}

\begin{acknowledgements}
We thank 
the referee, Andrei Tokovinin,
for helpful advice and
Jesper Storm for discussion.
This research made use of the VizieR catalogue access tool,
the SIMBAD database,
and the cross-match service provided by CDS Strasbourg, France,
he WFCAM Science Archive providing UKIDSS, the VISTA Science Archive,
the NASA/IPAC Infrared Science Archive (IRSA), operated by the Jet Propulsion
Laboratory, California Institute of Technology, under contract with the
National Aeronautics and Space Administration,
the SDSS DR12 Science Archive Server, as well as the DENIS and SSS images.
\end{acknowledgements}

\begin{appendix}

\section{Previously known CPM companion candidates confirmed with new data}
\label{Sect_confirmcpm}

%
%
\begin{table*}[t!]
\caption{Data on confirmed CPM companions of nearby F stars (for notes and references see Table~\ref{Tab_newcpmdata})}
\label{Tab_confirmedcpmdata}
\centering
\fontsize{7pt}{0.90\baselineskip}\selectfont
\begin{tabular}{@{}lrrc@{}c@{}cccccr@{}}     
\hline\hline
Name & RA (J2000) & DEC (J2000) & Epoch  & $\pi$ (Sep, PA)    & $\mu_{\alpha}\cos{\delta}$ & $\mu_{\delta}$ & N\_epochs & $J$ (2MASS) &SpT & Ref \\
HD     & [degrees]  & [degrees]   & [year] & [mas] ([arcsec,$^{\circ}$])& [mas/yr]                   & [mas/yr]      & & [mag]       & \\
\hline
1352A   & 4.458214   & $+$16.331064 & 1998.737 & 22.31$\pm$0.60 & $+$223.0$\pm$0.4 & $-$29.3$\pm$0.4 & & 6.29$\pm$0.02  & F6V & 2,3,4,S \\
1352B\tablefootmark{k,l} (?)  & 4.453571   & $+$16.327650 & 1998.737 & (20.2, 233)         & $+$225$\pm$4\tablefootmark{c}     & $-$41$\pm$4\tablefootmark{c}     & 41011100110 & 11.66$\pm$0.02 & $\approx$M4 &  2,2,5,5 \\
\hline
9826A\tablefootmark{$\bullet$}   &  24.199358 & $+$41.405582 & 1998.838 & 74.12$\pm$0.19 & $-$173.6$\pm$0.5 & $-$381.8$\pm$0.5 & & 3.18$\pm$0.21 & F9V & 2,3,4,S  \\
9826D\tablefootmark{k,l,m,s,y} & 24.210084 & $+$41.392387 & 1998.838 & (55.6, 149) & $-$182$\pm$4\tablefootmark{c} & $-$385$\pm$4\tablefootmark{c} & 41010000110 & 9.39$\pm$0.02 & $\approx$M4.5\tablefootmark{y} & 2,2,5,5 \\
\hline
18900A   & 45.898458  & $+$36.442005  & 1998.792 & 23.71$\pm$0.66 & $+$166.0$\pm$0.4 & $-$46.7$\pm$0.4 & & 6.55$\pm$0.03 & F8 & 2,3,4,S  \\
18900C\tablefootmark{k,l,s}   & 45.901000\tablefootmark{a}  & $+$36.441689\tablefootmark{a}  & 1998.792 & (7.4, 99)\tablefootmark{a}          & $+$156$\pm$20\tablefootmark{a,d}    & $-$87$\pm$20\tablefootmark{a,d}    & 01000100000 & $\lessapprox$12.5\tablefootmark{e} & $\lessapprox$M5 & 2,2,5,5 \\
\hline
31975A   & 73.273518 & $-$72.407852 & 1998.921 & 30.82$\pm$0.28 & $-$45.3$\pm$1.3 & $+$270.5$\pm$1.5 &  & 5.33$\pm$0.02 & F9VFe-0.5 & 2,3,8,S \\
31975B\tablefootmark{k,l} (?)  & 73.260386 & $-$72.405548 & 1998.921 & (16.5, 300)         & $-$51$\pm$18\tablefootmark{c}    & $+$248$\pm$8\tablefootmark{c}     & 31200000000 & 11.73$\pm$0.06\tablefootmark{g} & $\approx$M5 & 2,2,5,5 \\
\hline
33632A   & 78.322777 & $+$37.337357 & 1998.822 & 38.29$\pm$0.55 & $-$146.0$\pm$0.4 & $-$137.5$\pm$0.4 & & 5.43$\pm$0.02 & F8V & 2,3,4,S \\
33632B\tablefootmark{k,l,s}   & 78.326897 & $+$37.346207 & 1998.822 & (34.0, 20)         & $-$145$\pm$1\tablefootmark{c,h}     & $-$139$\pm$1\tablefootmark{c,h}     & 11010011010 & 10.38$\pm$0.03 & $\approx$M4 & 2,2,5,5 \\
\hline
49933A   & 102.7076176 & $-$0.5413573 & 2010.123 & 33.69$\pm$0.42 & $+$22.0$\pm$0.4 & $-$187.0$\pm$0.4 & & 4.99$\pm$0.02\tablefootmark{f} & F3V & 6,3,4,S \\
49933B\tablefootmark{l,m}   & 102.7083317 & $-$0.5397087 & 2010.123   & (6.5, 23)          & $+$7$\pm$15\tablefootmark{h}   & $-$199$\pm$18\tablefootmark{h}    & 01000012010 & 8.39$\pm$0.04 & $\approx$M0/sdM? & 6,6,5,5 \\
\hline
75289A\tablefootmark{$\bullet$}   & 131.918255 & $-$41.736645 & 1999.159 & 34.31$\pm$0.32 & $-$19.5$\pm$0.8 & $-$227.7$\pm$0.8 & & 5.35$\pm$0.02 & F9VFe+0.3 & 2,3,8,S \\
75289B\tablefootmark{k,l,n,o}   & 131.926087 & $-$41.735394 & 1999.159 & (21.5, 78)         & $-$4$\pm$13\tablefootmark{c}     & $-$212$\pm$15\tablefootmark{c}    & 31110000100 & 11.75$\pm$0.04 & $\approx$M5 & 2,2,5,5 \\
\hline
84999A   & 147.747331 & $+$59.038689 & 1999.885 & 28.06$\pm$0.20 & $-$294.7$\pm$0.4 & $-$150.7$\pm$0.4 & & 3.27$\pm$0.23 & F2IV & 2,3,4,S \\
84999B\tablefootmark{l,m}   & 147.741606 & $+$59.040123 & 1999.885 & (11.8, 296)         & $-$293$\pm$2     & $-$149$\pm$2     & 01000100010 & 8.78$\pm$0.03 & $\approx$M0 & 2,2,5,5 \\
\hline
89125A & 154.310830 & $+$23.106316 & 1998.071 & 43.85$\pm$0.36 & $-$412.9$\pm$0.4 & $-$96.2$\pm$0.4 & & 5.00$\pm$0.26\tablefootmark{f} & F6V & 2,3,4,S \\
89125B\tablefootmark{k,l,m,z} & 154.308779 & $+$23.107351 & 1998.071 & (7.7, 299) & $-$398$\pm$15\tablefootmark{d,z} & $-$118$\pm$15\tablefootmark{d,z} & 01000200000 & 8.36$\pm$0.03 & $\approx$M1.5\tablefootmark{z} & 2,2,5,5 \\
\hline
89744A\tablefootmark{$\bullet$}  & 155.544047 & $+$41.229637 & 1998.260 & 25.36$\pm$0.31 & $-$118.8$\pm$0.3\tablefootmark{j} & $-$139.1$\pm$0.3\tablefootmark{j} &  & 4.86$\pm$0.02 & F7V & 2,3,4,S \\
89744B\tablefootmark{k,l,m,s,u}  & 155.562043 & $+$41.240746 & 1998.260 & (63.0, 51)         & $-$123$\pm$5\tablefootmark{d,h}     & $-$131$\pm$6\tablefootmark{d,h}     & 01000100210 & 14.90$\pm$0.04 & $\approx$L0.5\tablefootmark{u} & 2,2,5,5 \\
\hline
102574A  & 177.0974417 & $-$10.3135272& 2014.258 & 23.23$\pm$0.45 & $-$105.2$\pm$0.7 & $-$107.6$\pm$0.7 & & 5.17$\pm$0.02 & F7V & 1,3,4,S  \\
102574D\tablefootmark{k,l,m,o,s}  & 177.0988817 & $-$10.3184158& 2014.302 & (18.3, 164)         & $-$108$\pm$7\tablefootmark{c,h}     & $-$109$\pm$7\tablefootmark{c,h}     & 11110000111 & 11.33$\pm$0.07\tablefootmark{g} &  $\approx$M4 & 1,1,5,5 \\
\hline
102634A & 177.2545422 & $-$0.3186442 & 2013.846 & 28.50$\pm$0.49 & $-$207.1$\pm$0.4\tablefootmark{j} & $+$5.4$\pm$0.4\tablefootmark{j} & & 5.21$\pm$0.02\tablefootmark{e} & F6V & 1,3,4,S \\
102634B\tablefootmark{k,l,o,r} (?) & 177.2605606 & $-$0.3233306 & 2013.731 & (27.5, 128)         & $-$201$\pm$2\tablefootmark{c,h,i}     & $+$2$\pm$2\tablefootmark{c,h,i}     & 41111302210 & 9.96$\pm$0.02 & $\approx$M3 & 1,1,5,5 \\
\hline
116457A & 201.308092 & $-$64.485161 & 2000.321 & 25.51$\pm$0.24 & $-$153.9$\pm$0.9 & $-$22.5$\pm$0.8 & &  4.94$\pm$0.24\tablefootmark{f} & F3IV & 2,3,8,S  \\
116457B\tablefootmark{s} & 201.304426 & $-$64.479156 & 2000.321 & (22.4, 345)         & $-$149.2$\pm$5.1\tablefootmark{i} & $-$16.3$\pm$5.5\tablefootmark{i} &            & 10.26$\pm$0.04 & $\approx$M3 & 2,2,10,5 \\
\hline
124553A & 213.588833 & $-$5.947687 & 1999.123 & 23.33$\pm$0.53 & $-$305.5$\pm$0.4 & $+$83.3$\pm$0.4 & & 5.32$\pm$0.02 & F9V & 2,3,4,S \\
124553B\tablefootmark{k,l} & 213.585935 & $-$5.952792 & 1999.123 & (21.1, 210) & $-$304$\pm$2\tablefootmark{c} & $+$84$\pm$3\tablefootmark{c} & 31010000111 & 10.53$\pm$0.02 & $\approx$M3  & 2,2,5,5 \\
\hline
126679A & 216.839157 & $-$14.838769 & 1998.603 & 22.42$\pm$0.93 & $-$172.3$\pm$0.8 & $-$59.7$\pm$0.8 & & 6.40$\pm$0.05 & F7V & 2,3,8,S \\
126679B\tablefootmark{l,$\ddagger$} & 216.848140 & $-$14.844262 & 1998.603 & (37.0, 122) & $-$173$\pm$6\tablefootmark{v}   & $-$66$\pm$4\tablefootmark{v}  & 31101000101 & 12.76$\pm$0.02 & $\approx$M5\tablefootmark{$\ddagger$} & 2,2,5,5 \\
\hline
129502A & 220.765132 & $-$5.658098 & 1999.132 & 54.73$\pm$0.20 & $+$108.9$\pm$0.4\tablefootmark{j} & $-$316.0$\pm$0.4\tablefootmark{j} & & 3.34$\pm$0.28\tablefootmark{f} & F2V & 2,3,4,S \\
129502B\tablefootmark{l,$\star$} & 220.754618 & $-$5.652752 & 1999.132 & (42.3, 297) & $+$104$\pm$4\tablefootmark{c,d}     & $-$319$\pm$4\tablefootmark{c,d}     & 21210100111 & 10.72$\pm$0.04 & $\approx$M5 & 2,2,5,5 \\
\hline
132052A & 224.295897 & $-$4.346479 & 1999.115 & 37.17$\pm$0.32 & $-$98.5$\pm$0.4 & $-$153.6$\pm$0.4 & & 4.13$\pm$0.28\tablefootmark{f} & F0V & 2,3,4,S \\
132052B\tablefootmark{r} & 224.290225 & $-$4.343605 & 1999.115 & (22.8, 297) & $-$92$\pm$14\tablefootmark{c,h} & $-$165$\pm$8\tablefootmark{c,h} & 11100000010 & 12.19$\pm$0.11\tablefootmark{g} & $\approx$M6 & 2,2,5,5 \\
\hline
132375A & 224.720010 & $-$4.989197 & 1999.115 & 29.61$\pm$0.47 & $-$360.4$\pm$0.4 & $-$105.0$\pm$0.4 & & 5.14$\pm$0.02\tablefootmark{f} & F8V & 2,3,4,S \\
132375B\tablefootmark{k,l,m} & 224.721083 & $-$4.986747 & 1999.115 & (9.6, 24)          & $-$378$\pm$21    & $-$95$\pm$12     & 01100000011 & 9.24$\pm$0.05 & $\approx$M1 & 2,2,5,5 \\
\hline
142529A  & 239.266144 & $-$48.162132 & 1998.447 & 20.64$\pm$0.48 & $-$101.1$\pm$1.0\tablefootmark{j} & $-$101.0$\pm$1.0\tablefootmark{j} & & 5.54$\pm$0.03 & F1IV & 2,3,8,S \\
142529B\tablefootmark{l,o}  & 239.255926 & $-$48.155563 & 1998.447 & (34.1, 314)         & $-$100$\pm$5\tablefootmark{c,i}     & $-$95$\pm$9\tablefootmark{c,i} & 41010000100 & 9.03$\pm$0.02 & $\approx$K7 & 2,2,5,5 \\
\hline
147449A & 245.518117 & $+$1.029096 & 2000.400 & 36.67$\pm$0.33 & $-$158.0$\pm$0.4 & $+$51.8$\pm$0.4 & & 4.29$\pm$0.22\tablefootmark{f} & F0V & 2,3,4,S \\
147449B\tablefootmark{l,w} (?) & 245.526848 & $+$1.021001 & 2000.400 & (42.9, 133) & $-$151$\pm$3\tablefootmark{c,d,h} & $+$47$\pm$3\tablefootmark{c,d,h} & 41110100110 & 8.68$\pm$0.03 & $\approx$M1.5\tablefootmark{w} & 2,2,5,5 \\
\hline
165670A & 271.711628 & $+$8.875950 & 2000.208 & 24.74$\pm$0.69 & $+$50.5$\pm$0.4\tablefootmark{j} & $-$147.2$\pm$0.4\tablefootmark{j} & & 6.00$\pm$0.02 &  F5V & 2,3,4,S \\
165670H\tablefootmark{l,s,$\otimes$} (??) & 271.711935 & $+$8.873803 & 2000.208 & (7.8, 172) & $+$86$\pm$7\tablefootmark{$\otimes$} & $-$119$\pm$7\tablefootmark{$\otimes$} & 01000000010 & 10.27$\pm$0.10 & $\approx$M3?/WD?\tablefootmark{$\otimes$} & 2,2,1,5 \\
\hline
166285A & 272.475023 & $+$3.119825 & 2000.419 & 21.31$\pm$0.31 & $+$15.9$\pm$0.4 & $-$193.1$\pm$0.4 & & 4.67$\pm$0.23 & F6V & 2,3,4,S \\
166285B\tablefootmark{k,l,s} & 272.474717 & $+$3.117978 & 2000.419 & (6.7, 189)          & $+$6$\pm$20     & $-$221$\pm$20    & 01000000010 & 8.26$\pm$0.13 & $\approx$K3 & 2,2,5,5 \\
\hline
176441A & 284.927192 & $+$16.252560 & 2000.208 & 21.38$\pm$0.83 & $-$99.1$\pm$0.4 & $-$121.4$\pm$0.4 & & 6.12$\pm$0.03 & F5 & 2,3,4,S \\
176441B\tablefootmark{l,r,s} & 284.924690 & $+$16.248047 & 2000.208 & (18.4, 208)         & $-$97$\pm$2\tablefootmark{c,h}     & $-$130$\pm$7\tablefootmark{c,h}     & 41001000110 & 12.10$\pm$0.04 & $\approx$M4 & 2,2,5,5 \\
\hline
185395AB & 294.1105231 & $+$50.2221189 & 2013.803 & 54.54$\pm$0.15 & $-$7.2$\pm$0.4 & $+$264.2$\pm$0.4 & & 3.88$\pm$0.28\tablefootmark{f} & F3V & 1,3,4,S  \\
185395E\tablefootmark{l,o,s,$\ast$}  & 294.0599208 & $+$50.2204692 & 2013.622 & (116.7, 267)        & $-$3$\pm$3\tablefootmark{c,i} & $+$261$\pm$8\tablefootmark{c,i} & 41010000110 & 8.98$\pm$0.03 & $\approx$M3.5\tablefootmark{$\ast$} & 1,1,5,5 \\
\hline
193307A & 305.420994 & $-$49.999325 & 1999.786 & 32.24$\pm$0.47 & $-$359,5$\pm$0.8 & $-$249.5$\pm$0.8 & & 5.24$\pm$0.02 & F9V & 2,3,8,S \\
193307B\tablefootmark{l,o} & 305.412983 & $-$49.996429 & 1999.786 & (21.3, 299)         & $-$348$\pm$7\tablefootmark{b,c,i}    & $-$262$\pm$12\tablefootmark{b,c,i}    & 31100000101 & 9.54$\pm$0.03 & $\approx$M2.5 & 2,2,5,5 \\
\hline
197373A & 310.074542 & $+$60.505283 & 1999.715 & 30.39$\pm$0.27 & $+$13.4$\pm$0.4 & $+$186.5$\pm$0.4 & & 5.12$\pm$0.04\tablefootmark{f} & F6IV & 2,3,4,S \\
197373B\tablefootmark{k,l} & 310.086529 & $+$60.508324 & 1999.715 & (23.9, 63)         & $+$14$\pm$2\tablefootmark{c,h,i} & $+$189$\pm$2\tablefootmark{c,h,i} & 31010000110 & 10.08$\pm$0.03 & $\approx$M3 & 2,2,5,5 \\
\hline
204485A & 322.034397 & $+$32.225300 & 1998.455 & 22.36$\pm$0.34 & $+$133.7$\pm$0.4 & $+$79.5$\pm$0.4 & & 5.16$\pm$0.05\tablefootmark{f} & F0V & 2,3,4,S \\
204485B\tablefootmark{r} (?) & 322.035177 & $+$32.228657 & 1998.455 & (12.3, 11)         & $+$117$\pm$6     & $+$69$\pm$6     & 01000000010 & 10.81$\pm$0.05 & $\approx$M3.5 & 2,2,1,5 \\
\hline
208502A & 328.748885 & $+$53.935604 & 2000.455 & 21.00$\pm$0.41 & $+$147.6$\pm$0.5 & $+$85.7$\pm$0.5 & & 5.99$\pm$0.02 & F5 & 2,3,4,S  \\
208502B\tablefootmark{l,r} (?) & 328.744662 & $+$53.937687 & 2000.455 & (11.7, 310)         & $+$163$\pm$6\tablefootmark{h}     & $+$88$\pm$5\tablefootmark{h}     & 01000011010 & 9.99$\pm$0.04 & $\approx$M3 & 2,2,5,5 \\
\hline
209268A & 330.896861 & $-$55.976936 & 1999.855 & 20.81$\pm$0.58 & $-$241.0$\pm$1.5 & $-$93.3$\pm$1.3 & & 5.85$\pm$0.02 &  F7V & 2,3,8,S \\
209268B\tablefootmark{l,m,o,$\dagger$} (?) & 330.892629 & $-$55.988174 & 1999.855 & (41.3, 192)\tablefootmark{x} & $-$255$\pm$5\tablefootmark{c}     & $-$84$\pm$6\tablefootmark{c}     & 41110000101 & 10.16$\pm$0.03 & $\approx$M2.5 & 2,2,5,5 \\
\hline
210855A & 332.9548478 & $+$56.8398414 & 2013.650 & 26.77$\pm$0.18 & $+$237.5$\pm$0.4 & $+$128.5$\pm$0.4 & & 4.68$\pm$0.29\tablefootmark{f} & F8V & 1,3,4,S  \\
210855C\tablefootmark{l,m,o,s} & 332.9762514 & $+$56.8301425 & 2013.357 & (54.7, 130)         & $+$238$\pm$1     & $+$130$\pm$1     & 10010013010 & 14.505$\pm$0.002\tablefootmark{g} & DA WD & 1,1,5,11 \\
\hline
221356A & 352.881212 & $-$4.087308 & 1998.723 & 38.29$\pm$0.54 & $+$178.0$\pm$0.9 & $-$191.9$\pm$0.9 & & 5.49$\pm$0.02 & F8V & 2,3,8,S \\
221356D\tablefootmark{l,s,t} (?) & 352.878967\tablefootmark{a} & $-$4.089861\tablefootmark{a} & 1998.723 & (12.2, 221)\tablefootmark{a}         & $+$158$\pm$9\tablefootmark{a,d}    & $-$202$\pm$14\tablefootmark{a,d}    & 01000100001 & $\approx$13.9\tablefootmark{e} & $\approx$L0.5\tablefootmark{t} & 2,2,5,5 \\
\hline
\end{tabular}
\end{table*}


In Table~\ref{Tab_confirmedcpmdata}, we
list F star companions known in the literature with
lacking or uncertain proper motion measurements.
In particular, we include companions previously listed
without proper motion in Tokovinin (\cite{tokovinin14})
or in the WDS catalogue (Mason et al.~\cite{mason01}).
For some URAT1 CPM companions recently identified by
Nicholson~(\cite{nicholson15}), we provide improved
proper motion and spectral type estimates.
CPM companions of Hipparcos stars already reported by
Gould \& Chanam{\'e}~(\cite{gould04}) and
L{\'e}pine \& Bongiorno~(\cite{lepine07})
are not considered here. Among the 31 confirmed
CPM companions, there are four within 25~pc (with
parallaxes of 40-75~mas).

\section{Rejected CPM companions}
\label{Sect_rejectcpm}


Table~\ref{Tab_rejectedcpmdata} lists
objects that appeared to be CPM companions according to SIMBAD,
but have newly determined by us proper motions in disagreement with
those of the F star primaries.

%
%
\begin{table*}[b]
\caption{Data on rejected CPM pairs including (nearby) F stars (for notes and references see Table~\ref{Tab_newcpmdata})}
\label{Tab_rejectedcpmdata}
\centering
\fontsize{7pt}{0.90\baselineskip}\selectfont
\begin{tabular}{@{}lrrc@{}c@{}cccccr@{}}     
\hline\hline
Name & RA (J2000) & DEC (J2000) & Epoch  & $\pi$ (Sep, PA)    & $\mu_{\alpha}\cos{\delta}$ & $\mu_{\delta}$ & N\_epochs & $J$ (2MASS) &SpT & Ref \\
     & [degrees]  & [degrees]   & [year] & [mas] ([arcsec,$^{\circ}$])& [mas/yr]                   & [mas/yr]      & & [mag]       & \\
\hline
Hip\,10529\tablefootmark{l,m,s,$\clubsuit$}    & 33.9222039  & $+$67.6770014 & 2013.577 & 4.6$\pm$18.3\tablefootmark{$\clubsuit$} & $-$7$\pm$6\tablefootmark{$\clubsuit$}       & $-$6$\pm$6\tablefootmark{$\clubsuit$}  & 01000000010 & 8.78$\pm$0.03 & F2  & 1,9,1,S \\
Hip\,10531    & 33.9324656  & $+$67.6711650 & 2013.591 & (25.3, 146)\tablefootmark{$\clubsuit$} & $+$518.1$\pm$0.3 & $-$304.8$\pm$0.3    &   & 5.66$\pm$0.03 & K2V & 1,1,4,S \\
\hline
HD\,187691A & 297.756919 & $+$10.415750 & 2000.573 & 52.11$\pm$11 & $+$242.3$\pm$0.3 & $-$136.5$\pm$0.2 & & 4.23$\pm$0.32\tablefootmark{f} & F8V & 2,3,3,S \\
HD\,187691B\tablefootmark{l,o,s,$\spadesuit$} & 297.752457 & $+$10.413567 & 2000.573 & (17.6, 244)\tablefootmark{$\spadesuit$}       & $-$6$\pm$22\tablefootmark{h,$\spadesuit$}      & $-$23$\pm$12\tablefootmark{h,$\spadesuit$} & 01000100010 & 10.31$\pm$0.04 & ?\tablefootmark{$\spadesuit$} & 2,2,5,5 \\
\hline
\end{tabular}
\end{table*}

\section{Selected finder charts}
\label{Sect_selfind}

\begin{figure}
   \centering
   \includegraphics[width=8.0cm]{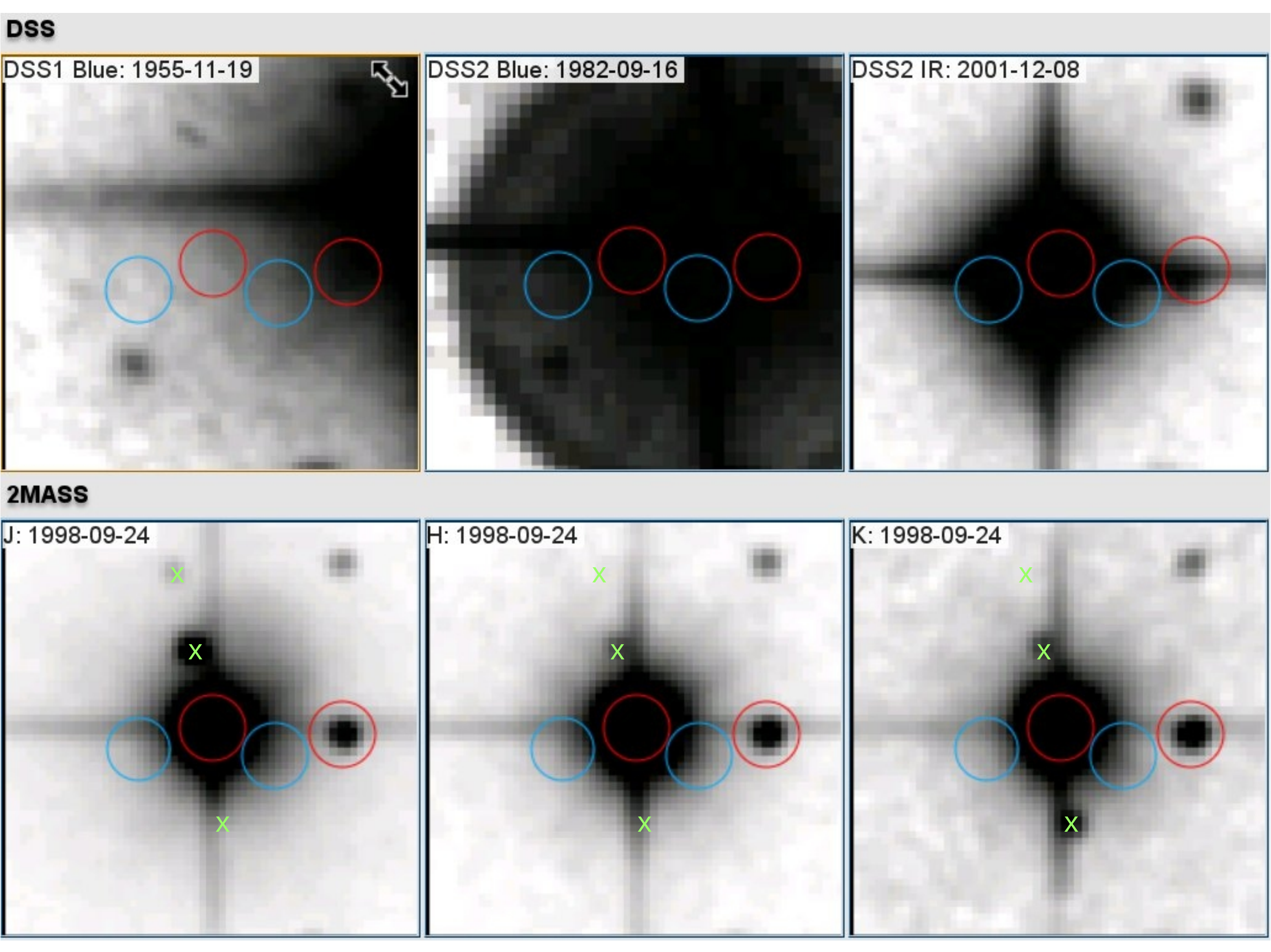}
      \caption{Finder charts of 1$\times$1\,arcmin$^2$
               (north is up, east to the left) from photographic
               plates of the Digitized Sky Surveys (DSS) and from
               2MASS of the CPM pair HD\,22879AB. Red circles:
               first-epoch (1998.732) positions (2MASS) of 
               components A (left) and B (right), blue circles:
               last-epoch (2012.955) positions (URAT1), 
               green crosses: 2MASS artefacts.
              }
         \label{F_HD22879AB}
   \end{figure}

Figure~\ref{F_HD22879AB} shows the 
IRSA finder charts of the new nearby ($d$$=$25.6~pc)
CPM pair HD\,22879AB. The faint $\approx$M6/sdM? 
companion HD\,22879B overlaps with the diffraction spike 
of its primary, the Gaia benchmark metal-poor star HD\,22879A,
but was visually measured in the 2MASS images
and the photographic IR image taken from the SSS.
The red and blue circles show the change in the positions of
the components from 1998 (2MASS) to 2013 (URAT1).

\begin{figure}
   \centering
   \includegraphics[width=11.0cm]{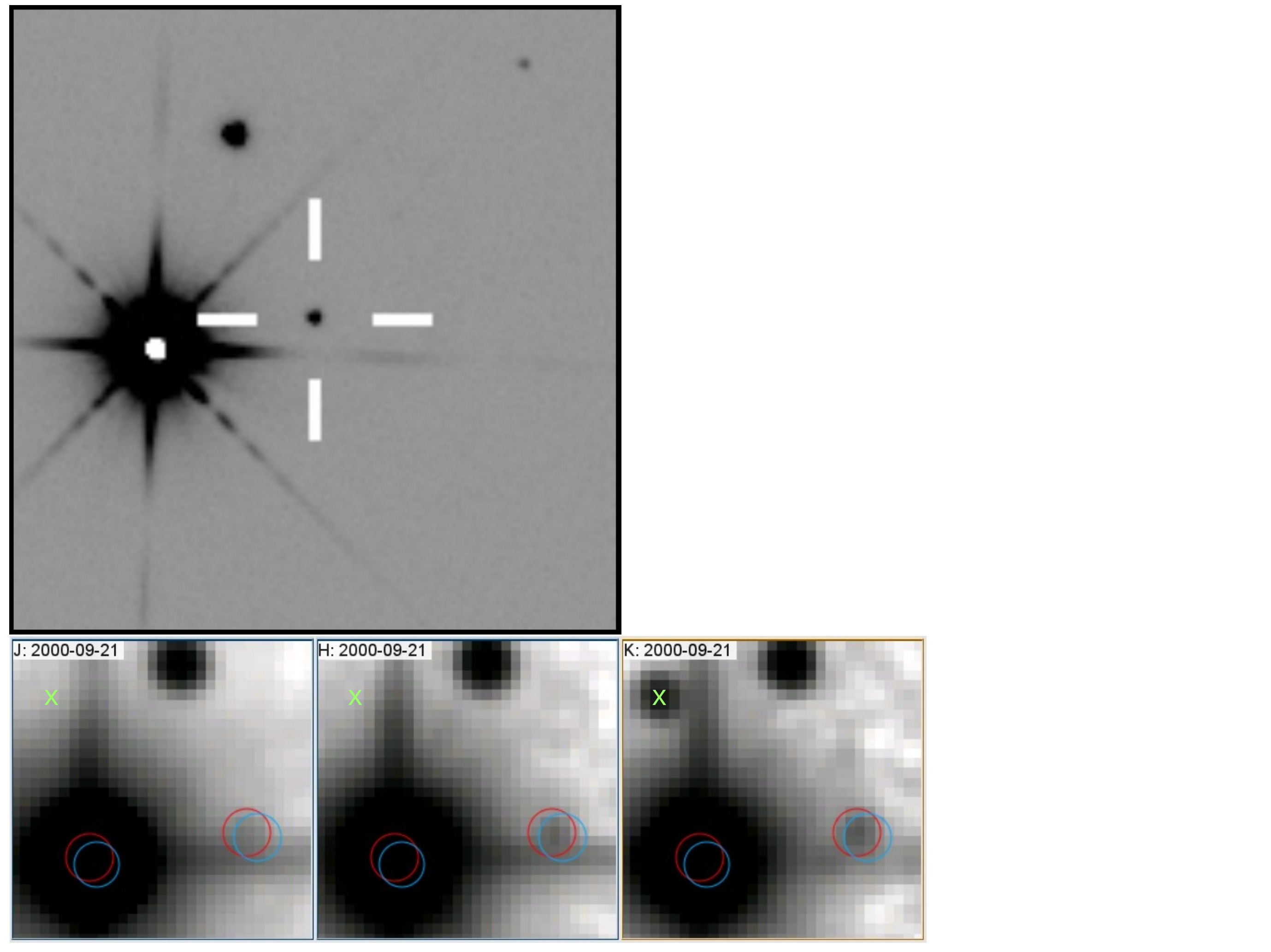}
      \caption{\textbf{Top:}
               Finder chart of 1$\times$1\,arcmin$^2$
               (north is up, east to the left) from 
               UKIDSS LAS ($K$-band) centred on HD\,3861B.
               \textbf{Bottom:}
               30$\times$30\,arcsec$^2$
               2MASS finder charts of the CPM pair HD\,3861AB. Red circles:
               first-epoch (2000.723) positions (2MASS) of
               components A (left) and B (right), blue circles:
               last-epoch (2010.592) positions (UKIDSS LAS),
               green crosses: 2MASS artefacts.
              }
         \label{F_HD3861AB}
   \end{figure}

As a second example, HD\,3861AB, our new CPM pair 
containing the latest-type ($\approx$L3.5) companion,
which was not detected on photographic plates, is shown in
Fig.~\ref{F_HD3861AB}. Again, the companion appeared close to 
the diffraction spike of its primary and was visually measured
in 2MASS. Whereas the small 2MASS finder charts illustrate
the proper motion of the components by the red and blue
circles, the UKIDSS LAS finder chart is centred on 
HD\,3861B.

\begin{figure}
   \centering
   \includegraphics[width=14.7cm]{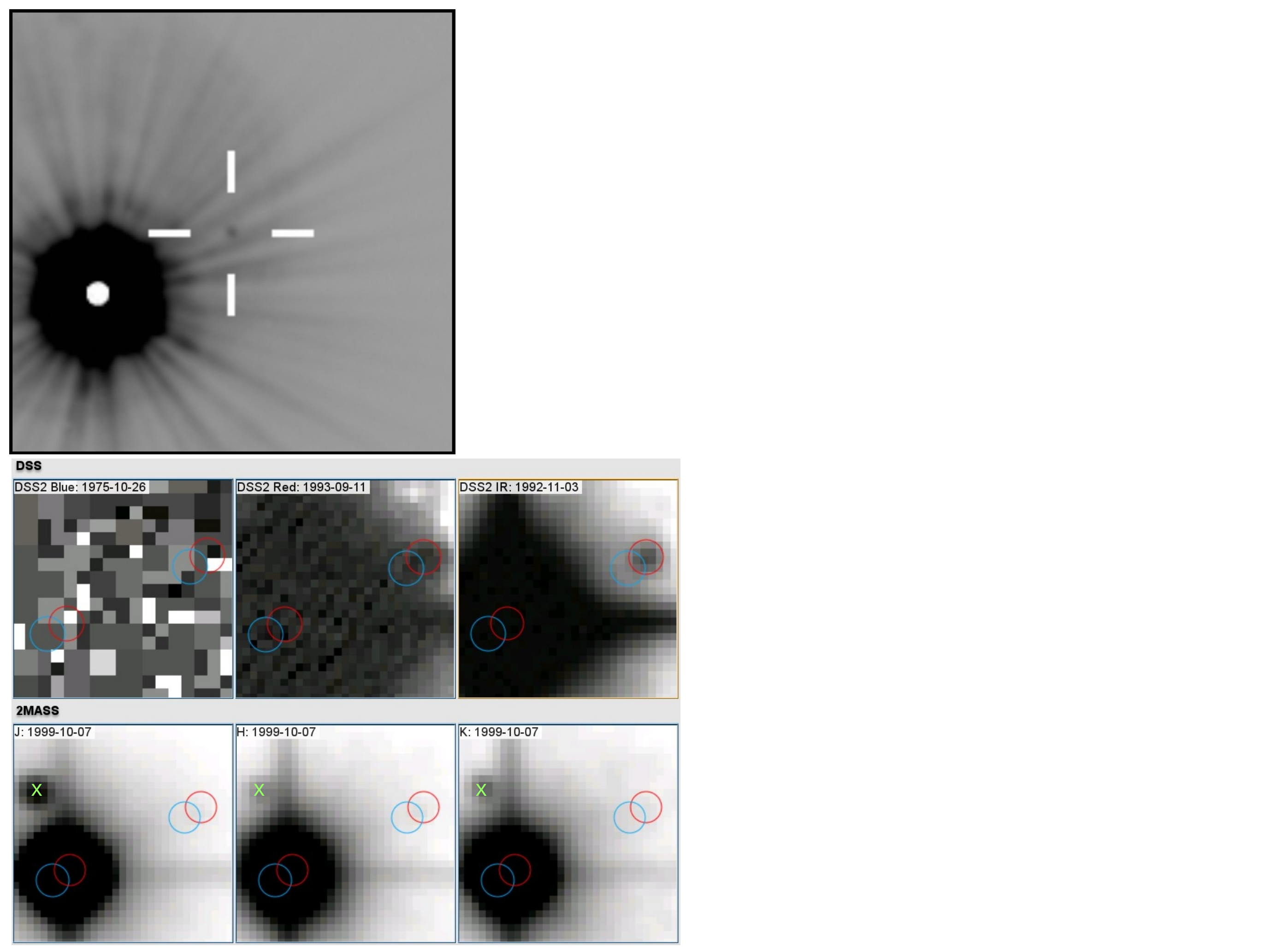}
      \caption{\textbf{Top:}
               Finder chart of 1$\times$1\,arcmin$^2$
               (north is up, east to the left) from
               VHS ($J$-band) centred on HD\,2726B.
               \textbf{Bottom:}
               Finder charts of 30$\times$30\,arcsec$^2$ from photographic
               plates of the Digitized Sky Surveys (DSS) and from
               2MASS of the CPM pair HD\,2726AB. Red circles: 
               first-epoch (1992.841) positions (DSS2 IR) of
               components A (left) and B ( right), blue circles:
               last-epoch (2011.625) positions (VHS),
               green crosses: 2MASS artefacts.
              }
         \label{F_HD2726AB}
   \end{figure}

Finally, we show the IRSA and VHS finder charts of 
HD\,2726AB in Fig.~\ref{F_HD2726AB}, where the new WD 
companion is clearly seen in
two of the photographic images, whereas it appears very
faint in 2MASS. Interestingly, the deeper near-infrared
data from the VHS provided the last-epoch data for this rather
blue object, thus helping us to confirm the CPM illustrated 
by the red and blue circles.

\end{appendix}

\end{document}